\documentclass[12pt]{article}
\usepackage[dvips]{color}
\usepackage{epsfig}
\usepackage{graphicx}

\begin{document}
\setcounter{page}{1}
\pagestyle{plain} \vspace{1cm}
\begin{center}
\Large{\bf Higher order quantum corrections of rotating BTZ black hole}\\
\small \vspace{1cm} {\bf B. Pourhassan \footnote{b.pourhassana@du.ac.ir}}\quad and \quad{\bf K. Kokabi
\footnote{kokabi@du.ac.ir}}\\
\vspace{0.5cm}$^{1,2}${\it School of Physics, Damghan University, Damghan, 3671641167, Iran}
\end{center}\vspace{1.2cm}
\begin{abstract}
In this work, we consider rotating BTZ black hole in three dimensions which is dual of one dimensional holographic
superconductors. We applied higher order corrections of the entropy, which interpreted as quantum corrections, to the thermodynamics quantities and study modified thermodynamics. We investigate stability of rotating BTZ black hole under effects of higher order quantum corrections, and find that they affect stability of black hole. So, the small black hole has some instabilities and critical points due to the quantum effects. We also study effect of correction terms on the dual picture of rotating BTZ black hole. 
\end{abstract}
\newpage
\section{Introduction}
Gauge/gravity correspondence is one of the important tools in fundamental physics. This correspondence, which is valid at the large N limit, is a relation between a weakly coupled theory of gravity in Anti-de Sitter (AdS) space-time and a strongly coupled conformal field theory (CFT) on the boundary. Dual picture of finite temperature CFT is presence of a black hole with the Hawking temperature in the bulk AdS space-time \cite{1,2,3,4,5}. In that case there are important application to the strongly coupled condensed matter systems, like construction of holographic superconductors and superfluids \cite{6,7,8,9}. In that case, relation between charged Reissner Nordstrom-AdS black holes in the presence of a charged scalar field and superconductors illustrated by the Ref. \cite{10}.
It is also possible to relate STU black holes \cite{11,12,13,14} to superconductor phase \cite{15}. Hence, study of black holes with dual CFT will be important from AdS/CFT point of view. There are several ways to study black holes. Among them, thermodynamics study is one of the hot and interesting topics of recent theoretical physics. In that case we know that thermodynamics of black objects modified due to thermal fluctuations coming from statistical fluctuations and interpreted as quantum corrections. Effects of such corrections calculated for the black holes entropy and found that at the leading-order it is logarithmic \cite{16, 17}. Hence, several black objects considered to find effect of logarithmic correction of entropy, for example in the Ref. \cite{18} the effects of the leading-order thermal fluctuations on a charged  black hole in anti de Sitter space has been studied. Also, corrected thermodynamics of G\"{o}del black hole due to this logarithmic correction investigated \cite{19} which may be applicable to the Kerr-G\"{o}del black hole statistics \cite{20}. Furthermore, stability of modified Hayward black hole investigated under effect of leading order corrections of entropy \cite{21}. Such studies are valid for all black objects, for example, such leading-order quantum correction have been applied to the thermodynamics of a black Saturns \cite{22, 23}. P-V criticality of several black holes investigated by corrected thermodynamics \cite{24, 25, 26, 27}. Such quantum gravitational effect has been studied recently by using Dumb holes \cite{28}. Quantum corrections to the Bekenstein-Hawking entropy of charged black hole using generalized uncertainty principle has been studied \cite{29}, which is possible to apply to Schwarzschild-Tangherlini black hole \cite{30}. Such quantum corrections will be important when the size of black hole be small due to the Hawking radiation. In that case thermodynamics of a sufficient small singly spinning Kerr-AdS black holes have been studied by the Ref. \cite{31}. As mentioned above, STU black hole is interesting from AdS/CFT correspondence point of view \cite{32}. In that case statistics of such black hole have been studied by the Ref. \cite{33}. Then effects of logarithmic correction in thermodynamics and hydrodynamics properties investigated \cite{34}.\\
Also, it is possible to calculate higher order corrections to the black hole entropy \cite{35}. Hence, in this paper we would like to consider rotating BTZ black hole \cite{36}, holographic dual of a one dimensional superconductor, and investigate the effects of higher order corrections of the entropy. We study corrected thermodynamics of rotating BTZ black hole. BTZ black holes \cite{37, 38} are of interesting kind of black holes. Hence, we would like to see effects of higher order correction of entropy on the thermodynamics quantities. This paper is organized as follows. In the next section we recall rotating BTZ black hole and review some important properties like horizon structure which is useful in the rest of paper. In section 3 we introduce origin of higher order corrections of the entropy, and in section 4 we obtain their effects on the thermodynamics of rotating BTZ black hole. In section 5 we discuss about P-V criticality and effects of correction terms on dual picture of black hole. Finally, in section 6 we give conclusion and propose several open problems for future works.
\section{Rotating BTZ black hole}
\begin{figure}[h!]
 \begin{center}$
 \begin{array}{cccc}
\includegraphics[width=75 mm]{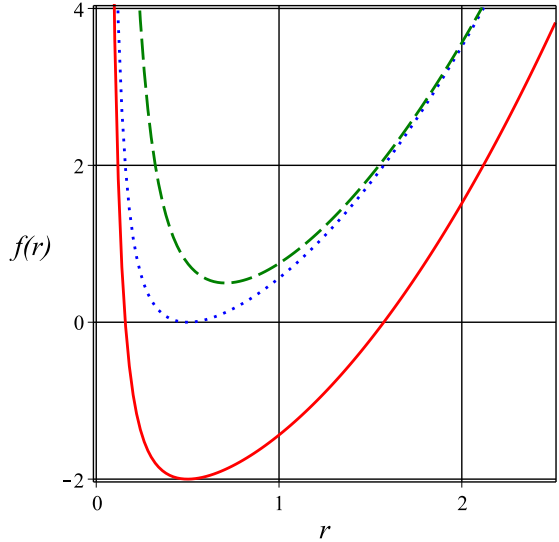}
 \end{array}$
 \end{center}
\caption{Horizon structure of rotating BTZ black hole.}
 \label{fig1}
\end{figure}

The rotating BTZ black hole embedded into the bulk which is described by a following Einstein-Maxwell action coupled to a charged scalar
field in 2+1 dimensions \cite{39},
\begin{equation}\label{0}
I=\int{d^{3}x\sqrt{-g}\left(R+\frac{2}{L^{2}}-\frac{1}{4}F_{\mu\nu}F^{\mu\nu}-|\nabla\Psi-iqA\Psi|^{2}-V(\Psi)\right)}.
\end{equation}
Three dimensional rotating BTZ black hole is given by the following metric \cite{36, 40},
\begin{equation}\label{1}
ds^{2}=-f(r)dt^{2}+\frac{dr^{2}}{f(r)}+r^{2}(d\phi-\frac{J}{2r^{2}}dt)^{2},
\end{equation}
where
\begin{equation}\label{2}
f(r)=-M+\frac{r^{2}}{l^{2}}+\frac{J^{2}}{4r^{2}},
\end{equation}
where $M$ and $J$ are constants which denote the black hole mass and angular momentum respectively. Also, in units where $G=\frac{1}{8}$ the length $l$ is related to the negative cosmological constant as $\Lambda=-\frac{1}{l^{2}}$.\\
Hawking temperature of black hole is given by,
\begin{equation}\label{3}
T=\frac{4r_{+}^{4}-J^{2}l^{2}}{8\pi l^{2}r_{+}^{3}},
\end{equation}
where
$r_{+}$ is event horizon radius, which given by,
\begin{equation}\label{4}
r_{\pm}^{2}=\frac{l}{4}(2Ml^{2}\pm2\sqrt{Ml^{2}-J^{2}}),
\end{equation}
which tell that $Ml\geq J$. It should be note that the coordinate $r$ is the direction into the bulk, while the boundary is parameterized
by the coordinates $t$ and $\phi$.\\
Also, the black hole mass defined as,
\begin{equation}\label{5}
M=\frac{r_{+}^{2}}{l^{2}}+\frac{J^{2}}{4r_{+}^{2}}.
\end{equation}
In the Fig. \ref{fig1} we draw $f(r)$ in terms of $r$ and show three different situations. Solid red line of the Fig. \ref{fig1} shows situation of $Ml>J$ where there are two distinct real roots $r_{\pm}$ where $r_{+}>r_{-}$. Dotted blue line represent extremal case of $Ml=J$ where $r_{+}=r_{-}$. Finally dashed green line denotes the case of $Ml<J$ which is naked singularity.

Also, Bekenstein-Hawking entropy of rotating BTZ black hole given by,
\begin{equation}\label{6}
S_{0}=\frac{\pi r_{+}}{2}.
\end{equation}
Finally, we can write the black hole volume as,
\begin{equation}\label{7}
V=\pi r_{+}^{2}.
\end{equation}

\section{Higher order correction of entropy}
Now, we would like to recall higher order corrected entropy and write its origin which is obtained by the Ref. \cite{35}. We work in the canonical ensemble of the given system include $N$ particle. In that case the partition function given by,
\begin{equation}\label{8}
Z=\int_{0}^{\infty}{\rho(E)e^{-\beta E}dE},
\end{equation}
where $\rho(E)$ is the quantum density of the system and defined as,
\begin{equation}\label{9}
\rho(E)=\sum_{n}{\Omega(E_{n})\delta(E-E_{n})},
\end{equation}
where $\Omega(E_{n})$ denotes number of microstates at energy $E_{n}$, and $E$ is the average energy of any given system. One can obtain canonical density of states by Laplace inversion of the equation (\ref{9}) as,
\begin{equation}\label{10}
\rho(E)=\frac{1}{2\pi i}\int_{-i\infty}^{i\infty}e^{\beta E+\ln{Z}}d\beta.
\end{equation}
The next step is usage of Taylor expansion of the entropy around equilibrium point $\beta_{0}$,
\begin{eqnarray}\label{11}
S&=&\beta E+\ln{Z}=S_{0}+\frac{1}{2!}\left(\frac{\partial^{2}}{\partial\beta^{2}}\ln{Z}\right)_{\beta_{0}}(\beta-\beta_{0})^2\nonumber\\
&+&\frac{1}{3!}\left(\frac{\partial^{3}}{\partial\beta^{3}}\ln{Z}\right)_{\beta_{0}}(\beta-\beta_{0})^3+\cdots.
\end{eqnarray}
where the second term of expansion vanishes due to having extremum at the equilibrium point $\beta_{0}$. Therefore, one can use $\beta-\beta_{0}=iy$ change of variable to write,
\begin{equation}\label{12}
\rho(E)=\frac{1}{2\pi}e^{S_{0}}\int_{-\infty}^{\infty}e^{-\frac{1}{2}\alpha_{2}y^{2}}\exp(\sum_{n=2}^{\infty}\frac{\alpha_{n}(iy)^{n}}{n!})dy,
\end{equation}
where
\begin{equation}\label{13}
\alpha_{n}\equiv\left(\frac{\partial^{n}}{\partial\beta^{n}}\ln{Z}\right)_{\beta_{0}},
\end{equation}
Now, for the large thermodynamics system one can use exponential expansion and solve integrals to obtain,
\begin{equation}\label{14}
\rho(E)=\frac{1}{2\pi}e^{S_{0}}\sqrt{\frac{2\pi}{\alpha_{2}}}\times \Theta,
\end{equation}
where
\begin{eqnarray}\label{15}
\Theta&\equiv&1+\sum_{n=2}^{\infty}\frac{\alpha_{2n}(-1)^n}{(2n)!!\alpha_{2}^{n}}\nonumber\\
&+&\frac{1}{2!}\sum_{n=3}^{\infty}\sum_{m=3}^{\infty}\frac{\alpha_{n}\alpha_{m}(-1)^{k}(2k-1)!!}{n!m!\alpha_{2}^{k}}+\cdots.
\end{eqnarray}
Then, one can use $\ln(1+x)\sim x$ approximation to obtain corrected entropy as following expression,
\begin{eqnarray}\label{16}
S&=&S_{0}-\ln{n}-\frac{1}{2}\ln{\alpha_{2}}\sum_{n=2}^{\infty}\frac{\alpha_{2n}(-1)^n}{(2n)!!\alpha_{2}^{n}}\nonumber\\
&+&\frac{1}{2!}\sum_{n=3}^{\infty}\sum_{m=3}^{\infty}\frac{\alpha_{n}\alpha_{m}(-1)^{k}(2k-1)!!}{n!m!\alpha_{2}^{k}}+\cdots.
\end{eqnarray}
We can see that leading-order correction is logarithmic while higher order corrections are also exist as a function of $S_{0}$. Then, following the Ref. \cite{35} one can write higher order corrected entropy as follow,
\begin{equation}\label{17}
S=S_{0}-\frac{1}{2}\ln{S_{0}T^{2}+\frac{\gamma}{S_{0}}},
\end{equation}
where $\gamma$ is a constant, for example $\gamma=-\frac{3}{16}$ for the ordinary BTZ black hole. It is also possible to fix coefficient by using thermodynamics properties.

\section{Modified thermodynamics}
In the case of infinitesimal $J$ one can obtain $S_{0}$ in terms of $T$ and hence giving inverse as follow,
\begin{equation}\label{18}
S_{0}=2\pi^{2}l^{2}T.
\end{equation}

\begin{figure}[h!]
 \begin{center}$
 \begin{array}{cccc}
\includegraphics[width=75 mm]{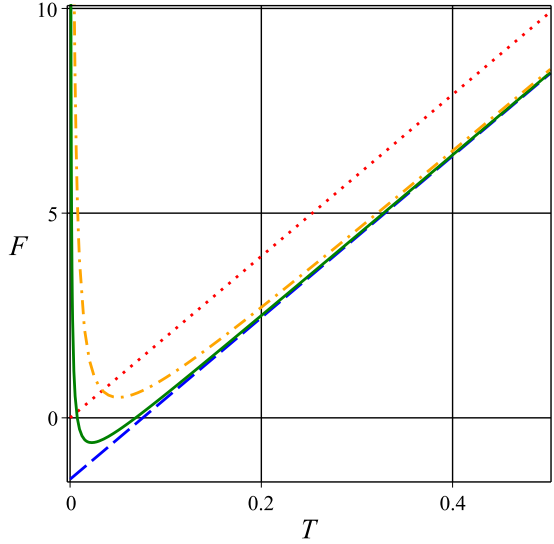}
 \end{array}$
 \end{center}
\caption{Helmholtz free energy in terms of temperature. Uncorrected (dotted red), Logarithmic corrected (dashed blue), Higher order corrected with $\gamma=-0.2$ (solid green), Higher order corrected with $\gamma=-1$ (dash dotted orange).}
 \label{fig2}
\end{figure}

Then, by using the Bekenstein-Hawking entropy (\ref{6}) and corrected entropy (\ref{17}) and the equation (\ref{18}) we have,
\begin{equation}\label{19}
S=2\pi^{2}l^{2}T-\frac{1}{2}\ln{2\pi^{2}l^{2}T^{3}+\frac{\gamma}{2\pi^{2}l^{2}T}}.
\end{equation}
Hence, we have,
\begin{equation}\label{20}
M=16\pi^{2}l^{2}T^{2}+\frac{J^{2}}{64\pi^{2}l^{4}T^{2}}.
\end{equation}
Also, angular velocity given by,
\begin{equation}\label{21}
\Phi=\frac{J}{2r_{+}^{2}}=\frac{J}{32\pi^{2}l^{4}T^{2}}.
\end{equation}
Helmholtz free energy obtained by using the following thermodynamics relation,
\begin{equation}\label{22}
F=-\int{SdT},
\end{equation}
which yields to the following relation,
\begin{equation}\label{23}
F=-\pi^{2}l^{2}T^{2}+T(\frac{\ln{2}}{2}+\ln{\pi}+\frac{\ln{l^{2}T^{3}}}{2}-\frac{3}{2})-\frac{\gamma \ln{T}}{2\pi^{2}l^{2}}.
\end{equation}
The second term of right hand side appear due to the logarithmic correction (leading-order) while the last term is due to the higher order correction. It is clear that effect of correction terms is infinitesimal and do not change general behavior of Helmholtz free energy, which has negative value. Also, it is clear that net value of the Helmholtz free energy is increasing function of the temperature.\\
Then, specific heat is given by the following relation,
\begin{equation}\label{24}
C=T(\frac{dS}{dT}),
\end{equation}
which yields to the following relation,
\begin{equation}\label{25}
C=2\pi^{2}l^{2}T-\frac{3}{2}-\frac{\gamma}{2\pi^{2}l^{2}T}.
\end{equation}
It is clear that the specific heat is completely positive quantity without correction of the entropy. However, presence of logarithmic correction makes some instabilities at low temperatures. On the other hand, higher order correction removes mentioned instabilities. All of these illustrated by the Fig. \ref{fig2}. In presence of higher order corrections, there is a minimum for the specific heat. If higher order coefficient $\gamma$ be sufficiently large, then the specific heat is completely positive and black hole is in stable phase. Otherwise for the small $\gamma$ we can see stable/unstable phase transition. Hence, we can find critical $\gamma$ where specific heat is zero and it is,
\begin{equation}\label{26}
\gamma_{c}=4\pi^{2}l^{4}T^{2}-3\pi^{2}l^{2}T.
\end{equation}

\begin{figure}[h!]
 \begin{center}$
 \begin{array}{cccc}
\includegraphics[width=75 mm]{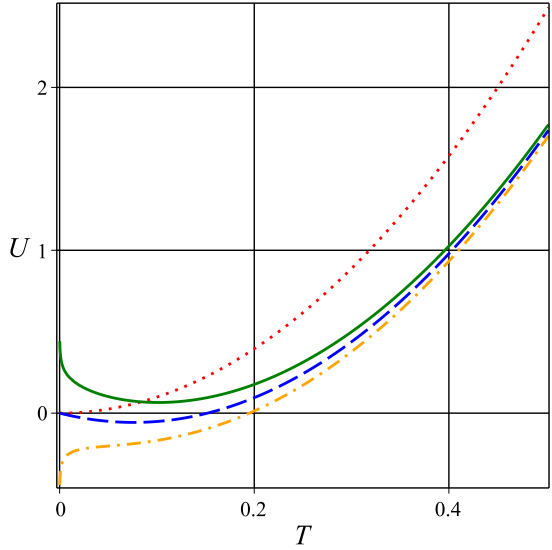}
 \end{array}$
 \end{center}
\caption{Internal energy in terms of temperature. Uncorrected (dotted red), Logarithmic corrected (dashed blue), Higher order corrected with $\gamma=1$ (solid green), Higher order corrected with $\gamma=-1$ (dash dotted orange).}
 \label{fig3}
\end{figure}

Now, by using the following thermodynamics relations,
\begin{equation}\label{27}
U=\int{CdT},
\end{equation}
or
\begin{equation}\label{28}
S=\frac{U-F}{T},
\end{equation}
one can obtain internal energy as follow,
\begin{equation}\label{29}
U=\pi^{2}l^{2}T^{2}-\frac{3}{2}T-\frac{\gamma\ln{T}}{2\pi^{2}l^{2}},
\end{equation}
Both relations (\ref{27}) and (\ref{28}) yields to the same result as expected. In the Fig. \ref{fig3} we draw internal energy in terms of the temperature and see effects of higher derivative corrections.

\section{Dual picture}
Because holographic superconductor is not depending on the entropy, hence we expect that correction terms of the entropy have no any important effect on the boundary field theory and superconducting phase transition. But it may be affect Van der Waals like behavior.
In order to discuss about dual picture, we need $P-V$ diagram to investigate possible Van der Waals behavior. Thermodynamics pressure given by the following relation,
\begin{equation}\label{30}
P=-\frac{\partial F}{\partial V}.
\end{equation}
It means that
\begin{equation}\label{31}
P=-{\frac {1}{16V\sqrt {{\frac {V}{\pi }}}{\pi }^{2}{l}^{2}} \left(  \left( -\pi \,V-4\,
\gamma \right) \sqrt {{\frac {V}{\pi }}}+  V \ln  \left( {\frac
{1}{32{\pi }{l}^{4}} \left( {\frac {V}{\pi }} \right) ^{3/2}}
 \right)  \right)}
\end{equation}

\begin{figure}[h!]
 \begin{center}$
 \begin{array}{cccc}
\includegraphics[width=75 mm]{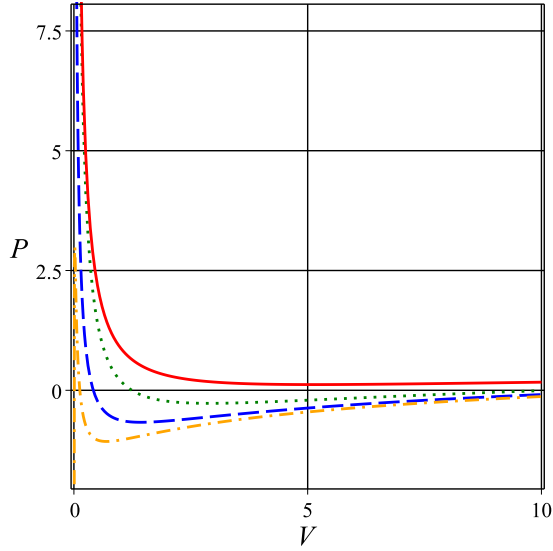}
 \end{array}$
 \end{center}
\caption{P-V diagram of the rotating BTZ black hole. $\gamma=0.6$ (solid red), $\gamma=0.2$ (dotted green), $\gamma=0$ (dashed blue), $\gamma=-0.2$ (dash dotted orange).}
 \label{fig4}
\end{figure}

So, in the Fig. \ref{fig4} we draw pressure in terms of the black hole volume to investigate possible critical point and Van der Waals behavior. We can see that, without correction terms, there is no critical point as well as phase transition. But in presence of higher order corrections we show that there is critical point depend on the higher order correction coefficient $\gamma$.

\begin{figure}[h!]
 \begin{center}$
 \begin{array}{cccc}
\includegraphics[width=75 mm]{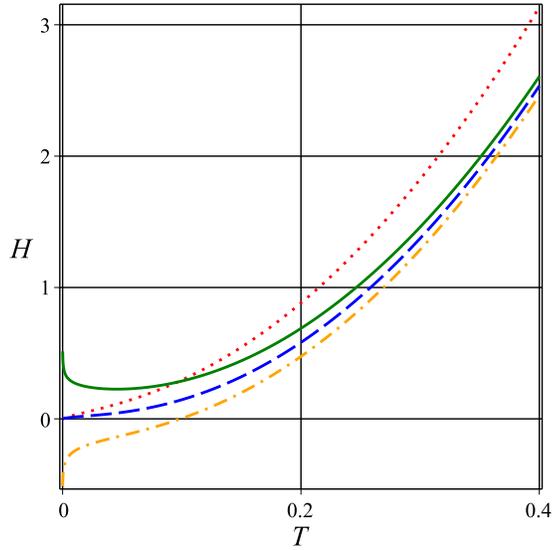}
 \end{array}$
 \end{center}
\caption{Enthalpy in terms of temperature. Uncorrected (dotted red), Logarithmic corrected (dashed blue), Higher order corrected with $\gamma=1$ (solid green), Higher order corrected with $\gamma=-1$ (dash dotted orange).}
 \label{fig5}
\end{figure}

Also, we can write pressure in terms of temperature as follow,
\begin{equation}\label{32}
P={\frac {\left( 2\,{\pi }^{2}T{l}^{2}-\ln
 \left( \sqrt {2}\pi  \right) -\frac{1}{2}\ln  \left( {T}^{3}{l}^{2}
 \right) +{\frac {\gamma}{2{\pi }^{2}T{l}^{2}}} \right)}{32{\pi }^{3}T{l}^{4}}  }.
\end{equation}
The last term is effect of higher order corrections which clearly increases (decreases) value of pressure if $\gamma$ be positive (negative).
Having temperature dependence pressure, we can extract enthalpy using the following relation
\begin{equation}\label{33}
H=U+PV,
\end{equation}
which yields,
\begin{equation}\label{34}
H={\frac {8\,{\pi }^{4}{l}^{4}{T}^{2}-6{\pi }^{2}T{l}^{2}-\ln
 \left( {T}^{3}{l}^{2} \right) T{\pi }^{2}{l}^{2}-2\,\ln  \left(
\sqrt {2}\pi  \right) T{\pi }^{2}{l}^{2}-2\,\gamma\,\ln  \left( T
 \right) +\gamma}{4{\pi }^{2}{l}^{2}}}
\end{equation}
In the Fig. \ref{fig5} we draw enthalpy in terms of temperature and see similar behavior with the internal energy. We can see that higher order corrected enthalpy with positive coefficient, as well as uncorrected case, is completely positive. However value of the enthalpy at low temperature is larger in presence of higher order correction. However increasing temperature reduces its value.\\

\begin{figure}[h!]
 \begin{center}$
 \begin{array}{cccc}
\includegraphics[width=75 mm]{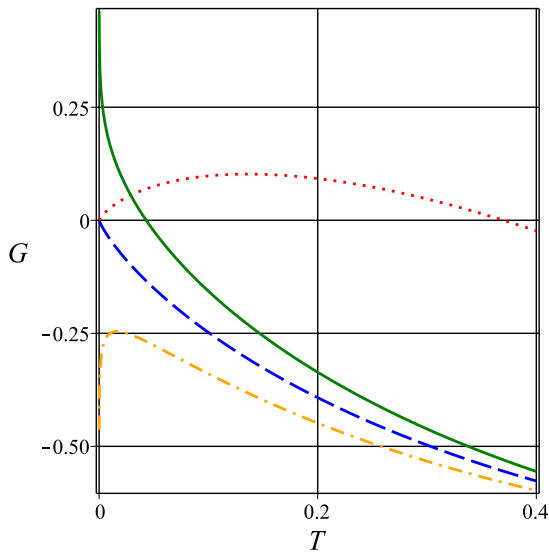}
 \end{array}$
 \end{center}
\caption{Gibbs free energy in terms of temperature. Uncorrected (dotted red), Logarithmic corrected (dashed blue), Higher order corrected with $\gamma=1$ (solid green), Higher order corrected with $\gamma=-1$ (dash dotted orange).}
 \label{fig6}
\end{figure}

Finally, one can obtain Gibbs free energy via,
\begin{equation}\label{35}
G=H-TS,
\end{equation}
which gives,
\begin{equation}\label{36}
G=-{\frac {2\,\ln  \left( \sqrt {2}\pi  \right) T{\pi }^{2}{l}^{2}+
T{\pi }^{2}{l}^{2}\ln  \left( {T}^{3}{l}^{2} \right) +2\,\gamma\,\ln
 \left( T \right) +\gamma}{4{\pi }^{2}{l}^{2}}}.
\end{equation}
In the Fig. \ref{fig6} we show that effect of corrected terms are crucial and important for the Gibbs free energy. Dotted line of the Fig. \ref{fig6} shows that uncorrected Gibbs free energy has a positive maximum at low temperatures. Then its value reduce and take negative magnitude. In presence of logarithmic correction, Gibbs free energy is completely negative variable. Including higher order corrections with positive coefficient case to initially positive Gibbs free energy which yields to negative value at high temperatures. Finally, higher order corrected Gibbs free energy with negative coefficient (dash dotted line of the Fig. \ref{fig6}) shows a maximum with negative value which yields to previous corrected case at high temperature.

\section{Conclusion}
In this paper, we considered rotating BTZ black
hole and investigated higher order quantum correction of the entropy. First of all we calculated thermodynamics quantities and discussed about horizon structure. Then, we considered thermal fluctuations of statistical physics, which could be interpreted as quantum corrections, and obtained higher order correction of the entropy which could be applied to several kinds of black objects. We have seen that the leading-order correction is logarithmic, while higher order correction is proportional to the inverse of entropy. We studied thermodynamics of the rotating BTZ black hole under effects of higher order corrections and found that Helmholtz free energy does not change general behavior. We analyzed specific heat and found that leading and higher order corrections are important in the stability of black hole. It means that when size of black hole become small due to Hawking radiation, it is no longer stable and quantum corrections yields to some instability of rotating BTZ black
hole. We also obtained effects of corrections on the internal energy, enthalpy and Gibbs free energy. Finally we found that in presence of logarithmic and higher order corrections, black hole behave like Van der Waals fluid and we can see critical point.\\
Here there are some kinds of black objects where one can apply higher order corrected entropy to study modified thermodynamics. For example one can consider Horava-Lifshitz black hole \cite{41} and see effects of higher order corrections. Schrodinger black holes \cite{42} as well as string black hole \cite{45} with hyperscaling violation is also interesting kind of black holes which may be considered for the higher order corrections. In the Ref. \cite{43} a new regular black hole considered and its statistics studied which may be extended as this paper. Also, Myerse-Perry black holes thermodynamics and statistics \cite {44} may be generalized to the case of higher order corrections.

\end{document}